# The fate of water in hydrogen-based iron oxide reduction

A.A. El-Zoka[a,b]*, L.T. Stephenson[a], S.–H. Kim[a], B. Gault[a,b], D. Raabe[a]*

[a] Max-Planck-Institut für Eisenforschung, Max-Planck-Strasse 1, 40237, Düsseldorf, Germany

[b] Department of Materials, Royal School of Mines, Imperial College, London SW7 2AZ, United Kingdom

* Corresponding Authors

## Abstract

Gas-solid reactions are cornerstones of many catalytic and redox processes that will underpin the energy and sustainability transition. The specific case of hydrogen-based iron oxide reduction is the foundation to render the global steel industry fossil-free, an essential target as iron production is the largest single industrial emitter of carbon dioxide. Our perception of gas-solid reactions has not only been limited by the availability of state-of-the-art techniques which can delve into the reacted solids in great structural and chemical detail, but we continue to miss an important reaction partner that defines the thermodynamics and kinetics of gas phase reactions: the gas molecules. In this investigation, we use the latest development in cryogenic atom probe tomography to study the quasi in-situ evolution of gas phase heavy water at iron-iron oxide interfaces resulting from the direct reduction of iron oxide by deuterium gas at 700°C. The findings provide new insights into the formation kinetics and location of water formed during hydrogen-based reduction of FeO, an its interaction with the ongoing redox reaction.

# Introduction

Iron is one of the most abundant elements in the earth's crust and in many stellar objects. It has several oxidation states, mainly Fe(II) and Fe(III), and crystallizes in different oxides such as hematite, magnetite, maghemite or wüstite. The reduction of iron oxides to iron, and the opposite oxidation, are both redox reactions of high relevance to the catalytic synthesis of ammonia[1] and hydrocarbons[2]; planetary morphogenesis[3]; magnets in structural biology[4]; biochemical reactions e.g. as in hemoglobin[5]; medical applications such as imaging[6], cell activation, cancer therapy[7]; metallurgy[8], as well as waste- and ground water treatment[9], to name but a few topics.

Here we focus specifically on the direct reduction of iron oxide by hydrogen, motivated by a recent surge in interest in moving away from the conventional carbon-based production of iron from its ores, that is among the largest single source of industrial carbon dioxide emissions responsible for climate change[10]. Many studies have addressed the thermodynamic and kinetic features of the underlying reactions [11–14], but little is known about the formation, location and role of the final reaction product, namely, water.

Addressing this fundamental question is essential for understanding the kinetics and achievable metallization of such reactions since water is expected to form at inner and outer surfaces, meaning that oxygen and hydrogen must first diffuse and then react. Yet, until now, information including where this recombination takes place, and in which state(s) the water forms, diffuses or gets stored remains unknown. Also, the role of water in porosity formation (because the oxide reduction is associated with large mass loss and gain in free volume in the solid) and as an interface layer which potentially blocks the further reduction is also not yet fully understood[11].

It is not only important to study how and where the water can form but also how it is removed from or stored at the reaction front. For better understanding of these storage and transport mechanisms it is relevant to know in which way the water resides on or in the solids and what are the associated diffusion mechanisms. The many lattice defects in the partially reduced solids, such as phase boundaries, grain boundaries, dislocations and vacancies, are likely also relevant for diffusion and nucleation of water. It is worth noting in that context, the presence of water can lead to undesired re-oxidation effects at the reaction front which again can have a high impact on the overall kinetics[15].

Tackling these questions requires high-resolution spatial probing of the reaction interfaces and their defect content, high analytical resolution, and real-space access to the reaction front and in-situ-like experimental conditions[16,17]. It must be also considered that for reduction of iron oxides with hydrogen (or other agents) is associated with significant volume changes between the adjacent phases and high mass loss, effects which cause high mechanical stresses, vacancies, nanopores, interfacial delamination as well as the formation of multiple types of lattice defects. This means that these redox phenomena are not just of a chemical but also of a structural and mechanical nature.

To study these effects, with particular attention to the location and dispersion of the water, we conducted here a quasi-in-situ near atomic scale real space study of the hydrogen based reduction of iron oxides using well controlled cryogenic and ultrahigh-vacuum workflow including the sample and reaction preparation and material probing using the latest development in atom probe tomography (APT)[18]. These features of the experimental workflow are important as water in its nanoscale form is elusive and can get lost in such experiments. Our preceding demonstrator experiments have indeed shown that APT analysis of water is possible[19,20].

Considering these constraints, we developed an experimental set-up consisting of a reaction module to enable the thermochemical treatments of already fabricated APT specimens. Following a first feasibility study reported in [18], we present here an investigation into the mechanism of iron oxide reduction by deuterium. We make the surprising observations that the reaction product is not formed (and removed) as a film on the reaction surface but stored in nanoscale dispersed form at the inner reaction regions and trapped as nanoscale water-metal inclusions inside the reduced iron.

## Results and Discussion

**Key Experiment and Oxygen Depleted Reaction Interface**

APT specimens are fabricated from single crystalline FeO (wüstite) samples along the crystallographic [100] direction via site-specific lift-out in a FEI Helios dual beam Xe-plasma focused ion beam (FIB) /scanning electron microscope (SEM). The APT specimens are then transported into an infrared laser reaction hub module using an ultra-high vacuum (UHV) transfer suitcase (Ferrovac VSN40S) maintained at cryogenic temperature (liquid nitrogen (LN2) cooled). In this coupled heating and reaction module different APT specimens are exposed to 50 mbar of deuterium gas ($D_2$) for 5, 10, 20, 30, and 60 s at 700°C. Before and after heating, samples are kept at the cryogenic temperature of 60 K inside this module. Mass spectrometry is used to verify that only deuterium is present in the reaction chamber. The so-reduced APT specimens are then transferred through the UHV suitcase to the atom probe at cryogenic -140 °C temperature. Details of the instruments and protocols are in the methods section.

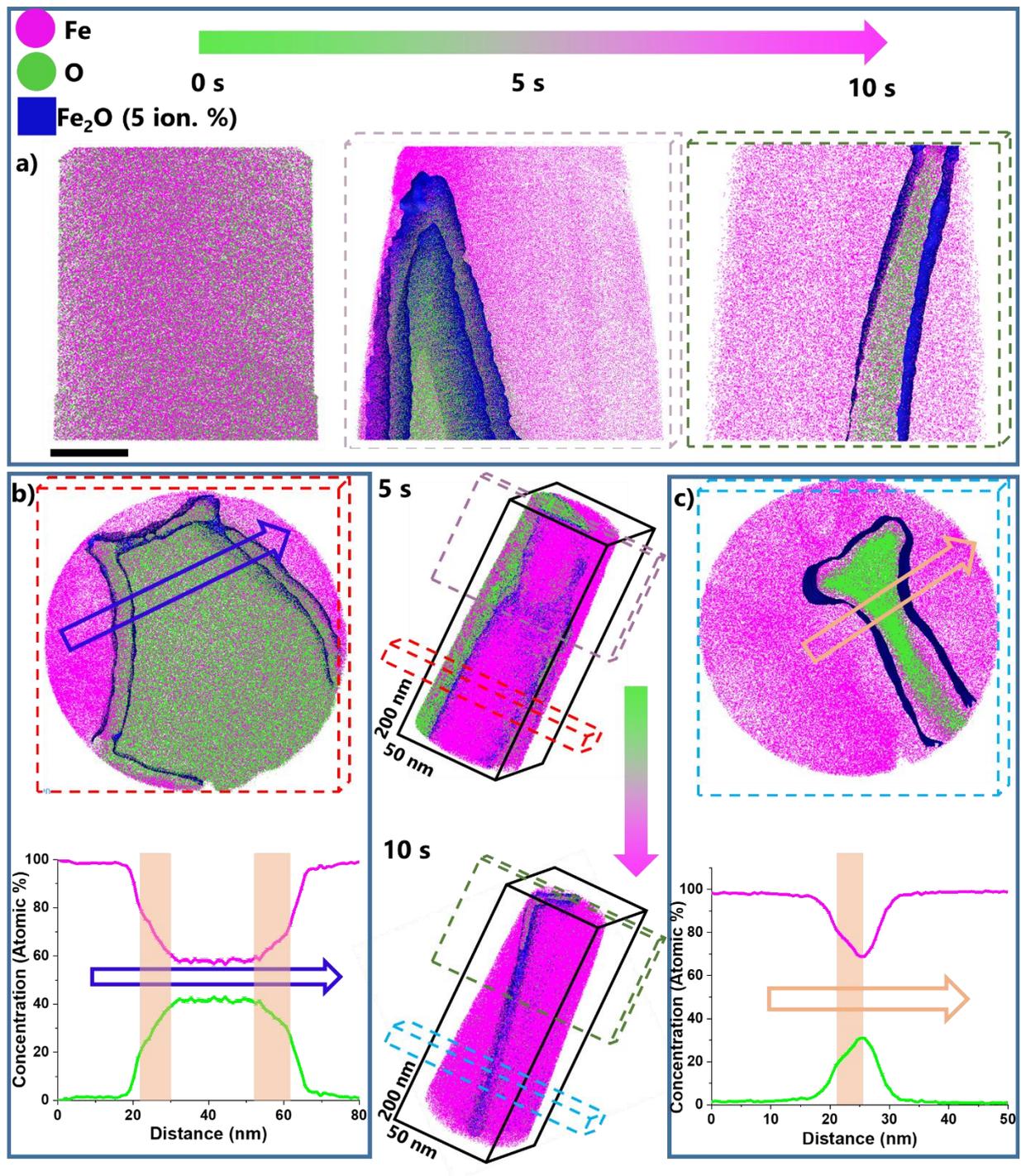

**Figure 1- Deuterium-based reduction of needle-shaped FeO specimens: a) APT maps of FeO during successive time steps of reduction at 700°C under 50 mbar deuterium gas atmosphere (D$_2$). b) Cross-section of FeO reduced for 5 s along with a compositional profile showing the Fe-O compositions along the reduction interface. c) Cross-section of FeO reduced for 10 s along with a compositional profile showing the Fe-O compositions along the reduction interface. The orange box-shaped shading highlights the oxygen-**

**depleted interface. Pristine FeO samples show a 52.8 ±0.0197 Fe-44.3 ±0.0196 O at. % composition. All error values are less than 1 at. %. The scale bar is 20 nm. Mass spectrum shown in Figure S1.**

APT analysis on pristine FeO (Figure 1 a) showed an overall composition of 52.8 ±0.0197 % Fe- 44.3 ±0.0196 at. % O. We also compare oxygen levels detected at different laser pulsing energies on the same pristine FeO sample, to examine the effect of field conditions on oxygen content detected. We find, by looking at Figure S2, that the oxygen content does vary with laser energy. Despite the fact that this FeO sample has a much lower impurity content compared to any mineral sample used in real applications, we do still notice several metallic impurities before reduction (including Ca, Al, and Na), as it is very hard to eliminate them completely. APT analysis of pure single crystal wüstite shows an under-stoichiometric oxygen content. This is possibly related to the recombination of oxygen ions at the atom probe tip surface into neutral species, depending on local electrical field strength[21]. This effect might lead to the apparent loss of oxygen as the spectrometry only counts charged particles. This was seen in previous APT study on wüstite reduction [22].

The reconstructed APT data show that after only 5-10s a considerable volume of the specimen has been reduced. Short term, interrupted reactions show the gradual increase in reduced Fe, and a corresponding decrease in pristine FeO and increase in Fe (**Figure 1**). The reduction interface had been "frozen in" at different times of the ongoing reaction. After 5 s of reduction at 700°C, regions of pure iron are formed at the surface of the specimen, albeit not covering it fully. The interior of the specimen is composed of an oxide region with an Fe-O (59 at. % Fe- 41 at. % O) composition similar to that measured in pristine FeO samples i.e. pure wüstite.

At the transition area between the reduced iron and the non-reduced oxide, an oxygen-depleted interface is already developed as clearly revealed in the chemical composition profiles (Figure 1b) (orange box) gathered across the reduction front and perpendicular to the tip direction (**Figure 1**). The thickness of this interface is found to be directly related to the progression of the reduction. More specific, the interfaces between iron and wüstite are characterized by a wide compositional Fe-O gradient, with a gradual drop in oxygen content from about 40 at. % inside the wüstite down to about 1 at. % inside the body-centred cubic α-Fe over a length of about 12-14 nm, which translates to a length of about 30 wüstite unit cells, across which the decay in oxygen content takes place. It is also worth noting the moderate change in the chemical gradient slope (areas in profile marked by orange boxes) when the oxygen content falls below about 18% (left hand side of **Figure 1b, bottom**) or, respectively, 30% (right hand side of **Figure 1b, bottom**). This change in slope might indicate (a) a change in oxygen diffusion mechanism, (b) a change in the remaining wüstite's structure features and/or (c) a change in the internal defect and / or pore structure which might alter the oxygen transport. In any case these data after 5s clearly show that the removal of oxygen from the wüstite is gradual, oxygen diffusion-mediated and sluggish process, without producing steep chemical gradients or any similar abrupt composition transitions. It is worth noting that Fe-oxides that contain oxide-stabilizing elements (such as Mg or Ti) would lead to more abrupt transitions between the surrounding iron and retained oxides[11].

Owing to the varying diameter of the APT specimens' needle shape, the reduced area cross-section varies along the shank. This means that the reduction happens faster at the top of the specimen, due to their higher surface-to-volume ratio. This is evident when comparing the thickness of these reduction interface shells at different distances from the initial specimen's apex, as the thickness of the iron grows from 5 to 10 nm (Figure S3).

The issue of distinguishing details of the oxygen content in heterogenous oxide structures by APT arises from the oxygen loss due to the effect of field applied during APT analysis[23,24]. This oxygen induced effect could introduce artefacts that mimic oxygen depleted zones in the sample. In our case we are able to confirm that the oxygen depleted interface is indeed due to the reduction process itself, by comparing the reduction at different areas of the reduced sample, as discussed above the interface's width changes with reduction progression. Also, the reduction interface we observe here is parallel to the specimen's main axis, whereas oxygen-movement due to applied field is usually driven by the field into the depth of the specimen (which was called field induced corrosion in the past[25]) and perpendicular to the specimen's main axis [26–28]. Finally, the variation across the reduction interface that we see in our analysis is more significant than that observed when analysing FeO at different laser energies (Figure S3). This increases our confidence in the oxygen depletion layer we present in Figure 1.

A similar analysis is obtained for the specimen reduced for 10 seconds (Figure 1c), where the tip is fully covered by pure Fe, with a remnant oxide in the centre of the tip. The Fe-O ratio in the remaining oxide region is at a maximum of 2.18. More specifically, the oxide composition within the span of 5 nm at the centre ranges between 10.6 at. % ± 0.48% and 31 at. % ± 0.31% of oxygen. This finding again shows that the oxygen depletion proceeds gradually. The data also show that the reduction leads to a nanoscale core-shell morphology, where the outbound diffusing oxygen that leaves the centre oxide must either (a) diffuse through the outer iron shell to the specimen's free surface to form deuterated water and / or (b) can form water also inside the specimen, provided that the oxygen mass loss has developed a nano-pore which can host that water. A clear distinction between the non-reduced oxide inside the specimen and the reduction interface becomes more difficult at these small dimensions, especially in the thinner region of the specimen, suggesting that the depleted oxygen interfaces and the already highly oxygen-depleted FeO layers overlap, leaving small oxide inclusions behind at advanced stages of reduction.

As an intermediate summary, the APT on-tip quasi in-situ deuterium reduction experiments show that the outbound diffusion of the oxygen creates a wide oxygen gradient inside the wüstite, extending over about 30 crystal cells. The actual removal mechanism of oxygen cannot be resolved at lattice scale, but it is likely that a change in diffusion mechanism of the oxygen inside the wüstite occurs when more and more oxygen vacancies are formed, probably leading to a transition of interstitial oxygen transport alone to interstitial plus substitutional oxygen transport. Deuterium transport is not considered to act as a bottleneck mechanism here, owing to its much higher diffusion coefficient, i.e. it may be assumed to be abundant throughout the APT specimen at the temperature and times probed here. An interesting feature is the change in the depletion slope, which might indicate a change in transport, structure or defects. Among these defects particularly the nano-pores, that evolve through the collapsing vacancies that the depleted oxygen leaves behind inside the wüstite, deserves particular attention as they might offer locations to host the water, as will be discussed below.

**Deuterium Accumulation at the Reduction Interface**

The rationale behind using deuterium instead of hydrogen is to differentiate the reactant in our specific reaction from hydrogen typically incorporated in analyses due to the Xe-plasma FIB preparation[29], or atom probe analysis [30,31]. In an ideal scenario, we would detect all of our reducing agent as $D_2^+$ at a mass signal of 4 Da to be able to fully distinguish D from H. However, due to the heterogeneity of the sample at these very early stages of the reduction, and the use of laser pulsing in the atom probe tomography experiment, detection of $D^+$ comes in peaks of 2 (primarily) and 3 Da (Figure S4), respectively. Theoretically, $H_2^+$ could also be detected at 2 Da, so in order to confirm that the high signals detected at 2 Da in our experiments can indeed be attributed to the D atom, we carried out a comparison of the ratio of 2 Da counts with respect to those for peaks at 1 and 3 Da for different specimens of Fe run under a range of electrostatic field conditions, indicated by the ratio of $Fe^{2+}$ to total Fe detected in each case, with and without deuterium exposure. As shown in Figure S5, the reduced samples in this study do indeed show higher intensity at 2 Da than usual, due to the introduction of the deuterium, confirming that the peak at 2 Da could be treated as a way of measuring the concentration of deuterium reacted during the reduction.

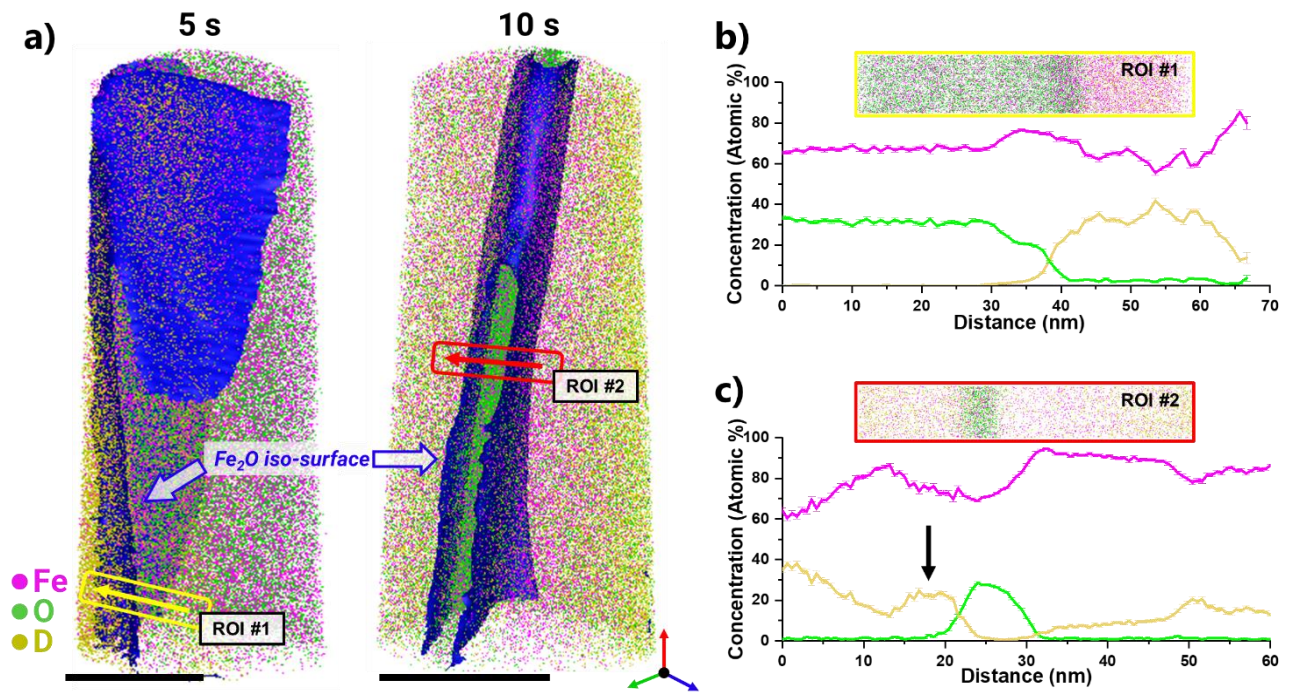

**Figure 2- Deuterium at the Reduction Interface: a) Composition of D$^+$ (2 Da) detected at different field conditions for different samples. b) Atom maps of samples in-situ reduced for 5 and 10 s at 700°C, respectively. c) Chemical profile across the reduction interface at 5 s. d) Chemical profile across the reduction interface at 10 s. Scale bars are 50 nm.**

The atom maps of Fe, O and D in Figure 2a and 2b reveal how the D tends to "attack" the oxide at selected areas, which favours a random depletion assumption as to where the near-surface reduction of iron oxides actually initiates. As shown in the 5s map (Figure 2a), the reduction front is not completely parallel to the reduction surface. Chemical profiling along one of those areas of D enrichment (Figure 2c), show that D injection could reach as high as 40 at. %, and that the D atom diffuses inwards, after the surface dissociation of the gaseous D$_2$.

After 10 s, the compositional profile of D has become more complicated, showing an enrichment right at the reduction interface. Accumulation of D is not seen at the same levels across the sample, confirming again that D reduction proceeds through random sites along the reduction interface at least near free surfaces.

Above all it is essential to use D instead of H for such experiments, to differentiate and track the intended reaction partners from the artificially acquired hydrogen from the environment and from sample preparation, an effect that can harm such delicate experiments, even under UHV conditions. The D sweeps the material very fast due to its high mobility. At least after

10 s at 700°C the D is essentially abundant throughout the material and does not act as a bottleneck to the overall redox reaction. D accumulates at the reaction interfaces / tip surfaces. It may be assumed - but cannot be resolved in the current experiments - that $D_2$ first dissociates at the free outer tip surface into 2 D atoms, and enters the material in single atomic form where it diffuses inbound, first following its chemical potential gradient and then saturating and partitioning among the iron, the wüstite, defects and the reaction interfaces.

A surprising observation is that the D is not detected within the oxide regions of the samples. Chemical profiles across the 10s reduced sample at multiple heights along the shank of the tip (Figure S6) confirm that D – when charged at the current pressure levels – does not appear into the oxide within the resolution limits of this experiment. The D signals at the centre of the oxide region are practically vanishing in all cases. Interestingly, a plot of the H detected at 1 Da also shows the same trend of partitioning (Figure S7), highlighting further that the reduction kinetics is controlled by oxygen diffusion to the reduction interface to react with D[32]. This behaviour could be specific to certain oxides, but not all of them, for example: a previous APT analysis on water-corroded Zircaloy shows affinity of H-related species to zirconium oxide[33].

Once again, the effect of electrostatic field conditions on H levels detected at 1 Da is investigated using a parameter sweep on FeO APT analysis to ensure minimal impact on our chemical profiles. Figure S8 shows that the variation in detected H is within 10 % between 60 and 100 pJ. H levels do increase more significantly at 20 pJ, which is probably due to an increase in specimen voltage to maintain the same rate of evaporation (detection rate)[34].

The reduction starts at the free wüstite surface of the specimens. The gradual depletion in oxygen in the immediate surface regions leads to a gradient in the chemical potential of the oxygen between the highly oxygen-depleted reaction surface / interface and the interior of the (gradually depleting) wüstite. This gradient in chemical potential leads to outbound diffusion of oxygen, through the newly formed iron layer. The two electrons from the anionic oxygen are taken up by the iron cations, rendering them metallic. These metallic iron atoms then relax and attach to the adjacent metallic iron layer. This elementary step leads to the further inward growth of the iron layer. At the beginning of the redox reaction, recombination between O and D inside of the solid into heavy water will likely not occur, due to the lack of any free volume. This is likely to change as the reaction proceeds further, continuously producing oxygen vacancies at the wüstite-iron interface region. These vacancies can condensate and collapse into nano-pores. These pores could host internally formed heavy water. This means that in principle the water can form not only at the nearest external surfaces but also inside of the pores in the solid. Details will be discussed in the next section.

**Where is the Deuterated Water?**

We expect the formation of $D_2O$ as a by-product of the reaction of D with FeO. In our work, and due to the fast freezing of samples after reduction, we are indeed able to observe a clear distinct signal at 20 Da indicating the formation of $D_2O$ as expected (Figure S9). However, due to the fact that calcium is also detected in the sample analysed prior to reduction, it may also be that the signal at 20 Da accounts for Ca. Deconvolution of peaks at 20-21-22 Da, is done using the APsuite software[35]. This confirmed that the isotope abundances do not completely match the measured intensity at 20 Da. Our analysis shows that ~1000 ions cannot be attributed to Ca and are instead probably due to $D_2O$. In agreement with ex-situ APT studies done before on reduced iron oxide ores[22,36] metallic impurities can partition or accumulate at the reduction interface. Only by using the deconvolution methods in this investigation we are able to catch the subtle presence of $D_2O$. After 5 seconds of reduction, an atomic scale layer of Ca and $D_2O$ is seen to form along the reduction interface as revealed in Figure 3a. By looking at the atom map for 10 s, we discover that this layer evolves into clusters (or nano-droplets) at the reduction interface, as shown in Figure 3a. Since it has been observed that the porosity formed in reduced Fe is due to the escape of water vapor[22], we believe that these nanoscale frozen water droplets form inside such pores that were observed in bulk samples to form during reduction[11].

While water is practically electrically non-conductive, previous studies have shown that water can be analysed by APT along with incorporated metallic ions[19]. In this case, water is rather confined, and it is expected that it would act as a dielectric, based on simulations[37]. Another important question is whether these nano-droplets or, respectively, layers of observed water are liquid or vaporized during the reduction, as the exact confinement and pressure conditions inside such pores are not yet fully understood. Recent studies have shown that confining water to a 2 - 4 nm could decrease its melting point by ~15 % [38]. But since the samples are quenched to 80 K, and are analysed at 50 K at the atom probe, we expect that the water is in a frozen solid state during the actual APT analysis.

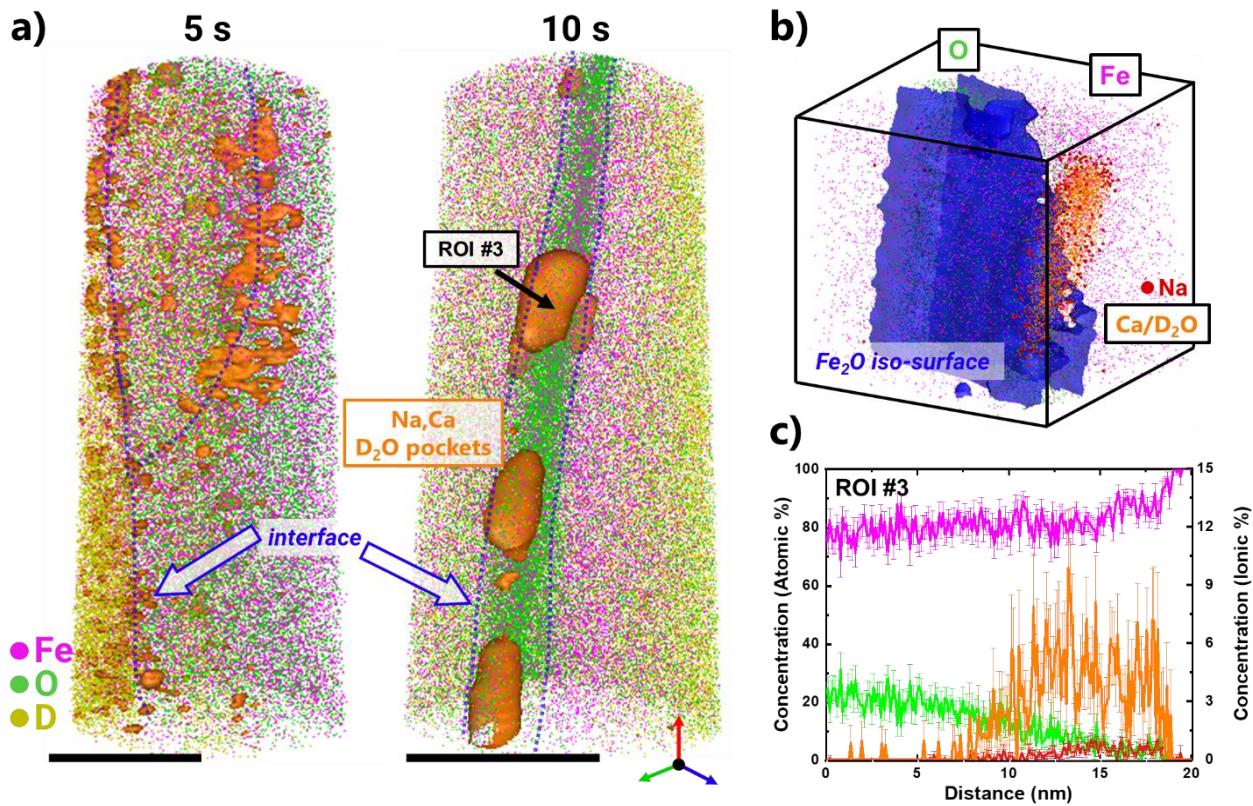

**Figure 3- Formation of heavy water droplets: a) Atom maps of FeO reduced for 5s and 10s, showing the formation and growth of D₂O clusters. b) A single water nano-droplet formed after 10 s at the FeO-Fe interface, along with a chemical profile across showing the high concentration of D₂O and Na. c) Atomic % concentration profiles across a nano-droplet for Fe, O and Na, and ionic % concentration profile for Ca-D₂O measured at 20 Da. Ca-D₂O isoconcentration surfaces are at the value of 4 at. %. Scale bars are 50 nm.**

By focusing on one of the 10 nm clusters formed, we notice further that the clusters of water tend to hold also enriched levels of other metallic impurities, mainly Na (7.14 at. ± % 2.2 %). These impurities also originate from the laboratory-grown oxide used in this study. This was found also in previous APT analysis on this same material[18]. Other impurities are also observed at the water cluster sites, such as Al and N (Figure S10). Previous work on nanoscale inclusions in geological samples also show that light metals tend to dissolve in water[39,40]. The impurities in this study are not likely to play an essential role in the reduction and water formation kinetics, as also seen before in studies by Kim et al. on natural ores [11]. Yet, we note that some impurities are included in these water clusters. The subsequent analyses of APT specimens reduced for longer times show that tips become fully reduced, while the water-impurity clusters grow much larger fully segregated to the surface of the tip at 60s (Figure S11).

The analysis of the heavy water in this section confirms the preliminary conclusions that we had drawn above, namely, that droplets of nano-confined water are formed at the reaction interfaces, in part even inside of the material, at the hetero - interfaces. This is attributed to the nano-porosity that might host that water. It should be also emphasised here that these mechanisms do not coincide but occur in sequence, i.e. first the D sweeps the material; second, the oxygen diffuses outbound, leaving behind vacancies and nano-pores; and third, water can form inside of these pores or at the free tip surfaces.

## Conclusions

Using in-situ high resolution chemical characterization by aid of a well-controlled cryogenic UVH workflow combining preparation, D-based reduction reaction and APT, the reduction of iron oxide by deuterium is now seen at unprecedented chemical and spatial resolution scales. A few main so far unknown atomic-scale characteristics of the reduction reaction were identified, namely, $D_2$ accumulation at the reaction interface; $D_2$ dissociation into D; immediate surface layer reduction and formation of heavy water with the oxygen from the wüstite; formation of a core-shell system with the newly formed iron as shell at the APT tip's free surface and the wüstite as the core phase; inbound diffusion of D through the iron layer and partitioning of D among all phases and defects, yet negligible D partitioning into the wüstite phase; gradual formation of an oxygen chemical potential gradient between any next free surface and the centre region of the partially and of the unreduced wüstite regions; outbound diffusion of oxygen through the wüstite and / or through the iron to the next free available inner or outer surface; formation of nano-pores from the oxygen vacancies; and the internal formation of heavy nano-water droplets at these nano-pores (some of which are connected to the surface while others are confined).

## Materials and Methods

### Single Crystalline FeO samples:

To start with a model specimen, a single crystal wüstite sample oriented towards the [100] direction, (an orientation accuracy of <0.1° to 0.05°) across its thickness is used. The sample is lab grown through the Czochralski (CZ) method, provided by Mateck GmbH. In the CZ method, the oxide is formed by inserting of a small seed crystal into an oxide melt in a crucible, pulling the seed upwards to obtain a single crystal[41]. Thus, eliminating factors such as impurities and porosity which might be found in ore pellets. The reduction protocol and preliminary obtained results are discussed in the following sections.

### FIB/PFIB Preparation

Specimens are mounted on a laser-ablated cold-rolled 304 stainless steel (304SS) TEM half-grid (sourced from JPT and CAMECA) following recently developed protocols[18].

### Atom probe tomography

Atom probe tomography (APT) analysis is performed on pristine FeO to measure the detected compositions of Fe in ratio to oxygen, and to measure impurity compositions. A parameter sweep was employed to compare the Fe/O ratios before and after reduction.

Cameca LEAP 5000 XR is used for all reduced APT tips. APT experiments for reduced FeO specimens are conducted in laser-pulsed mode, with laser energy of 50 pJ and 70 pJ, pulse frequency of 200 kHz, set point temperature of 45K and 1% detection rate. Data reconstruction and analysis is carried out using AP Suite 6.0.

APT specimens in the direction perpendicular to the thickness of the FeO substrate [100] direction are prepared via standard site-specific lift out procedure[42] using a FEI Helios dual beam Xe-plasma FIB/SEM. UHV carry transfer suitcases (Ferrovac VSN40S) are employed[43] for the current study. Cryogenic temperatures are maintained inside the suitcases by liquid nitrogen. The suitcases are designed to hold a modified Cameca APT puck and has a 50-cm long wobblestick which ends with a PEEK-insulated puck manipulator. A pressure of $10^{-10}$ mbar can be achieved inside suitcases. The suitcase could be mounted onto the experimental platforms through specially designed load locks (Ferrovac VSCT40 fast pump-down docks), pumped via a 80 L/s turbopump (Pfeiffer HiPace 80).

**Reaction Hub Measurements**

In line with previously reported protocols, IR laser in the reaction hub module is used to heat samples at the following times: 5, 10, 20, 30, 60, 120 seconds. All of these heating treatments are performed in an environment of 50 mbar of deuterium. Before and after heating, samples are kept at a cryogenic temperatures of 60 K using the cryo-chiller. Mass-spectrometry below is used to verify that only deuterium is present in the reaction chamber, as the presence of hydrogen could complicate analysis, and presence of water vapor could cause condensation on APT tips which was proven to deter reaction. Sample is then transferred through UHV suitcase to the atom probe.

## Acknowledgements

We thank U. Tezins, C. Broß, and A. Sturm for support to the FIB and APT facilities at MPIE. Funding: We are grateful for the financial support from the BMBF via the project UGSLIT and the Max-Planck Gesellschaft via the Laplace Project. A.A.E.-Z., L.T.S., S.-H.K., and B.G. acknowledge financial support from the ERC-CoG-SHINE-771602.

## Competing Interests

The authors declare that they have no competing interests.

# Supplementary Information

**The fate of water in hydrogen-based iron oxide reduction**

**A.A. El-Zoka \*, L.T. Stephenson, S.–H. Kim, B. Gault, D. Raabe\***

\* Corresponding Authors

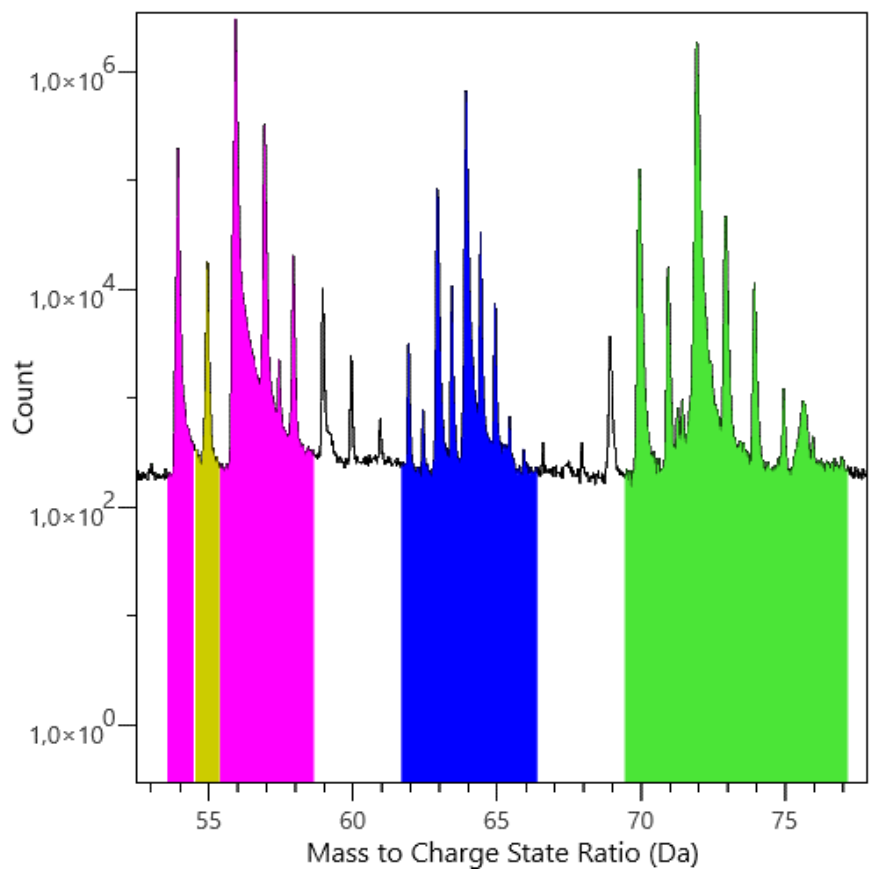

Figure S1. A partial mass-spectrum for the 10s data, showing, the detected Fe (purple), $Fe_2O$ (blue), and FeO (green).

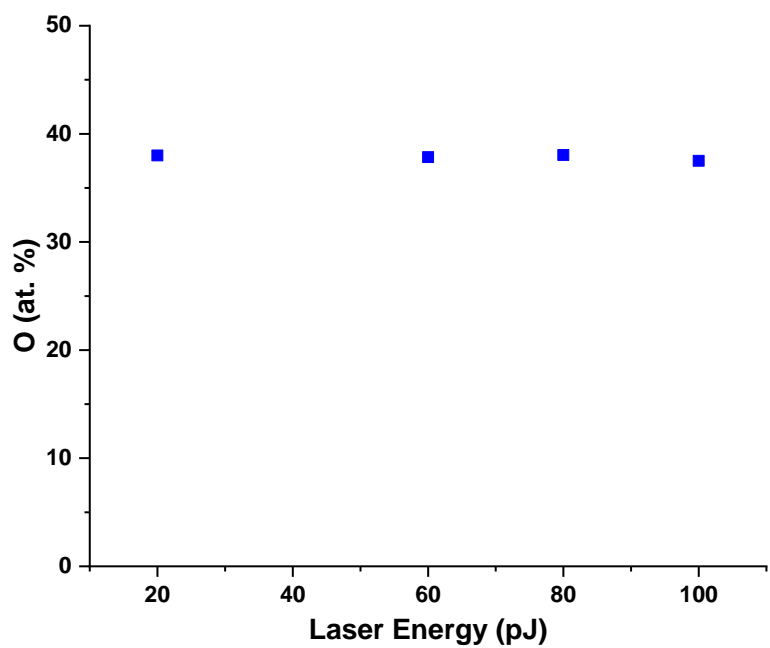

**Fig S2. Oxygen composition, detected at different pulsing energies.**

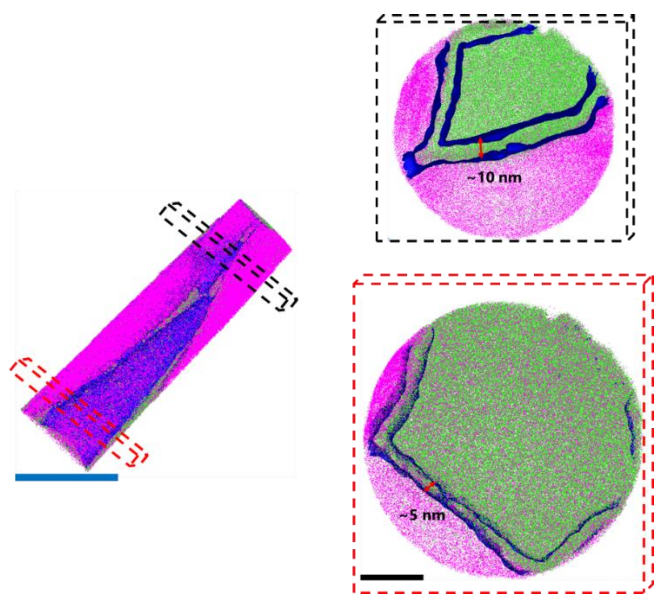

**Figure S3. Effect of tip radius on the local progression of reduction. Comparison of reduction interfaces at two different lengths along the tip.**

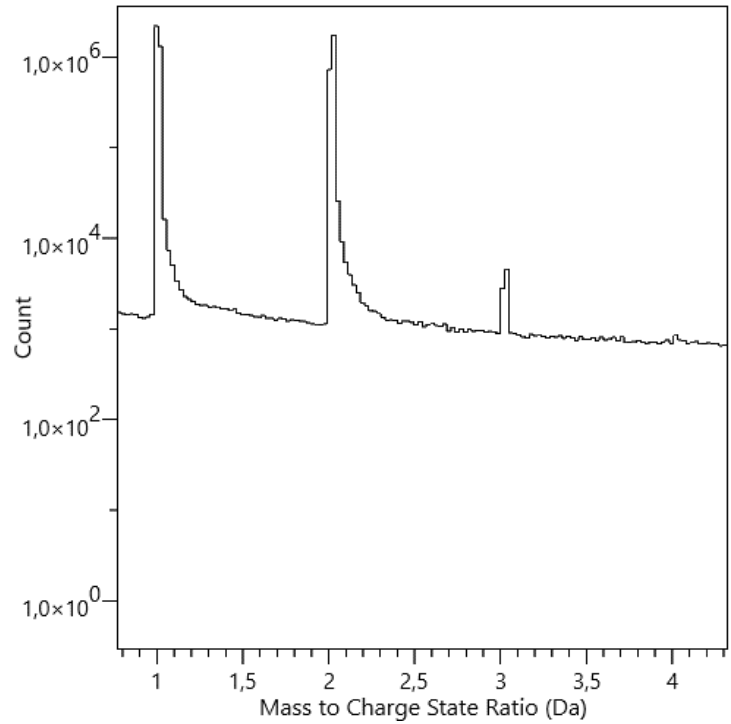

**Figure S4.** A partial mass-spectrum for the 10s data, showing, peaks at 1 Da (H), 2 Da ($H_2$ or D), and 3 Da (DH).

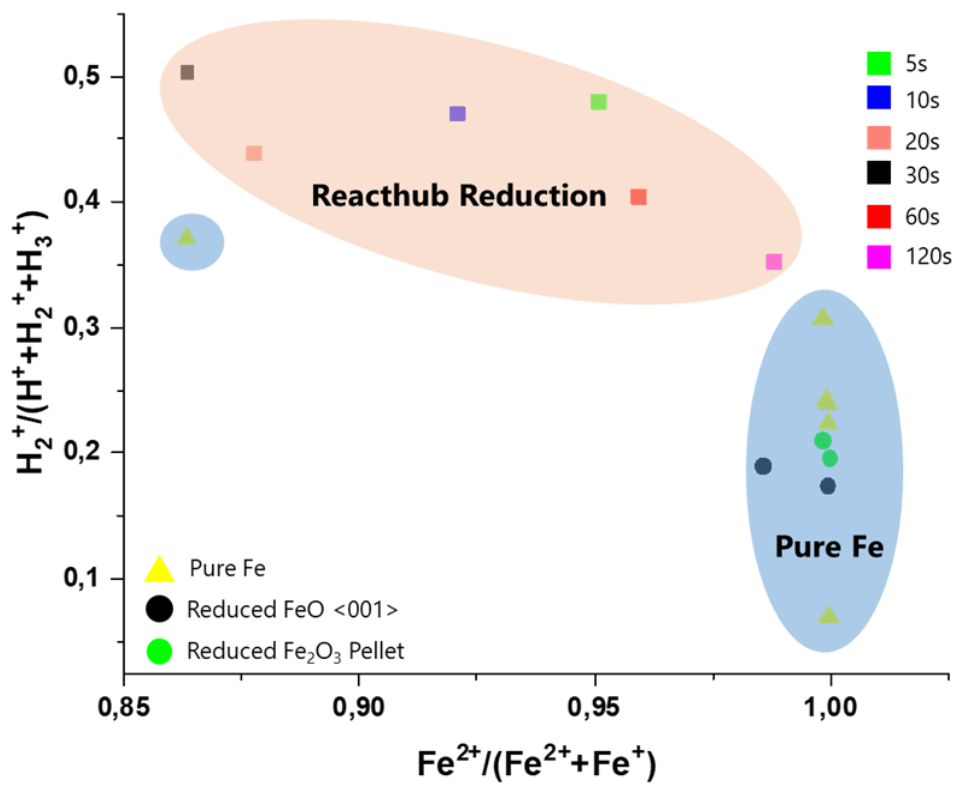

**Figure S5.** A comparison of peak count ratio for H₂ between samples reduced and frozen, and others including pure untreated Fe, FeO reduced and transferred at room temperature, and natural Fe₂O₃ reduced and transferred at room temperatures. All frozen deuterated samples show a clear consistent trend of having higher than expected counts for the peak at 2 Da.

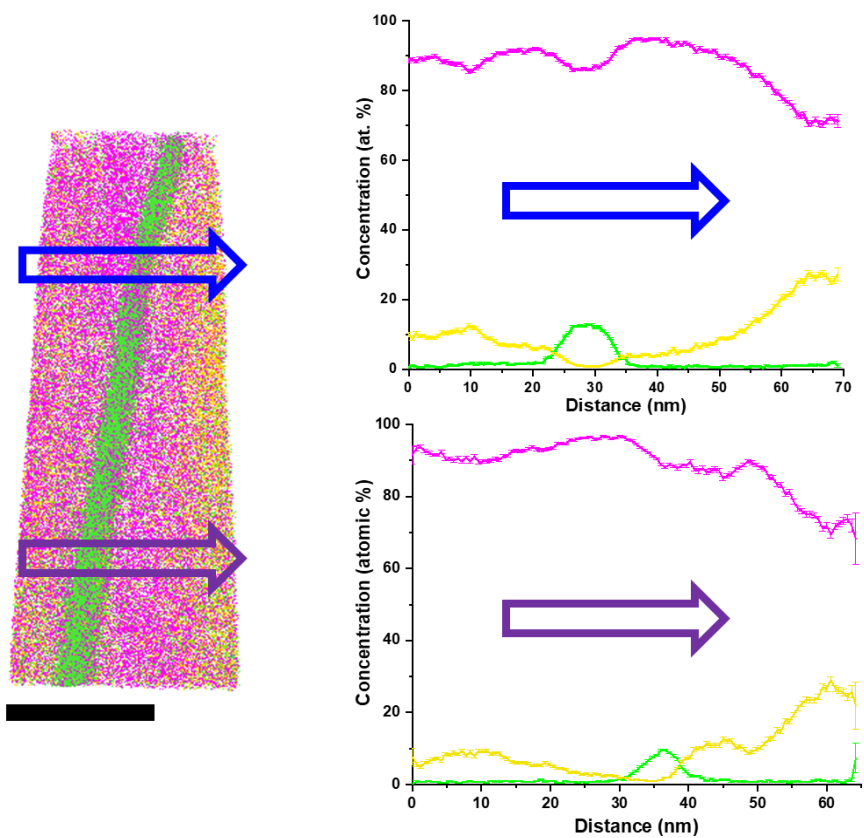

**Figure S6.** Chemical profile across the reduction interface at 10 s along different locations in the APT sample. Scale bar is 50 nm.

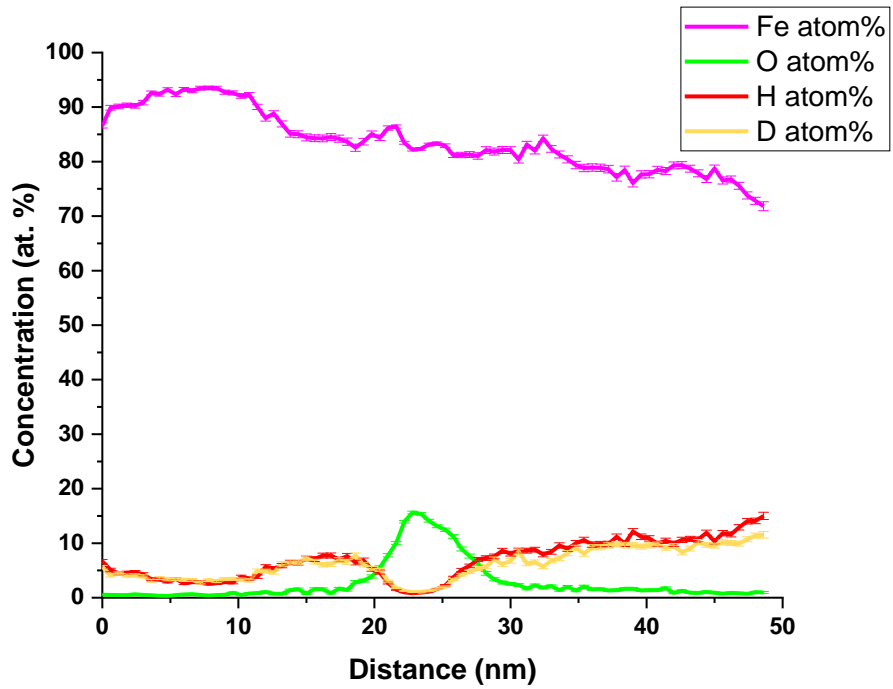

**Fig S7. Oxygen composition, detected at different pulsing energies.**

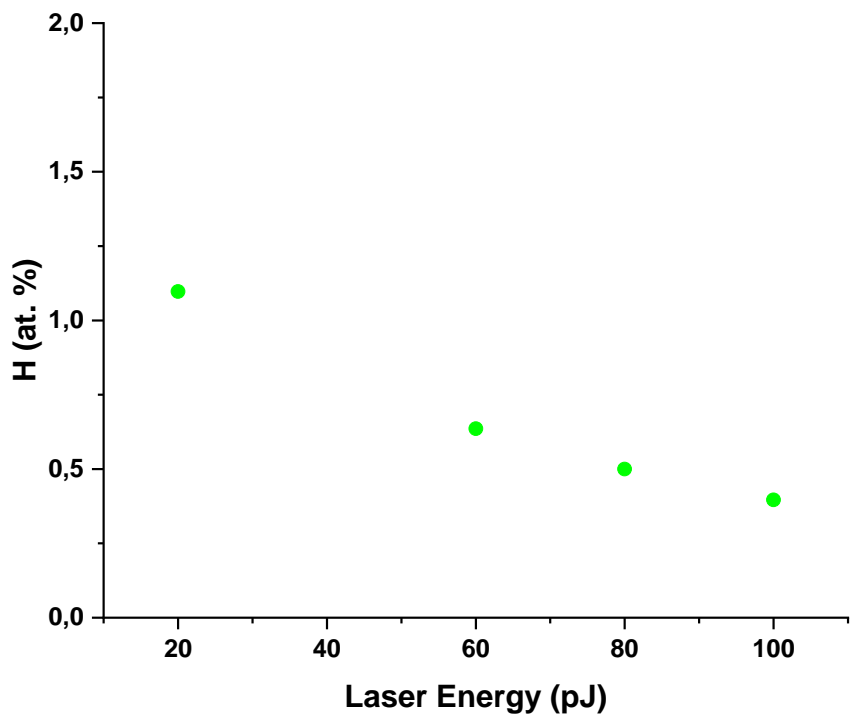

**Fig S8. H at. % composition, detected at different pulsing energies.**

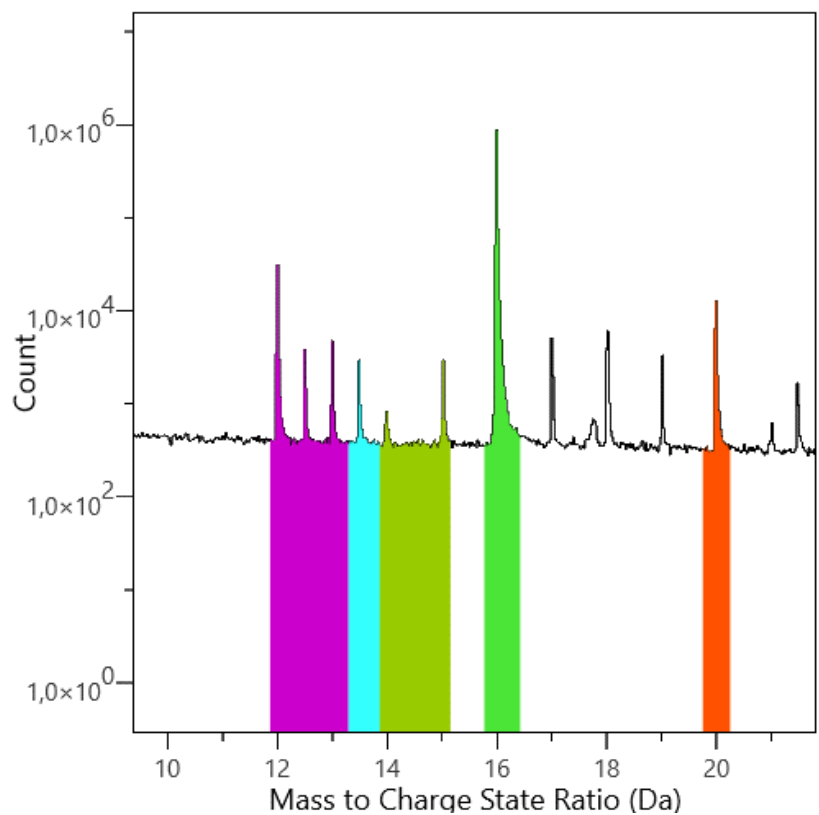

**Figure S9.** A partial mass-spectrum for the 10s data, showing signals detected for Mg (purple), Al (light blue), N (light green), O (green), and $D_2O$ (red).

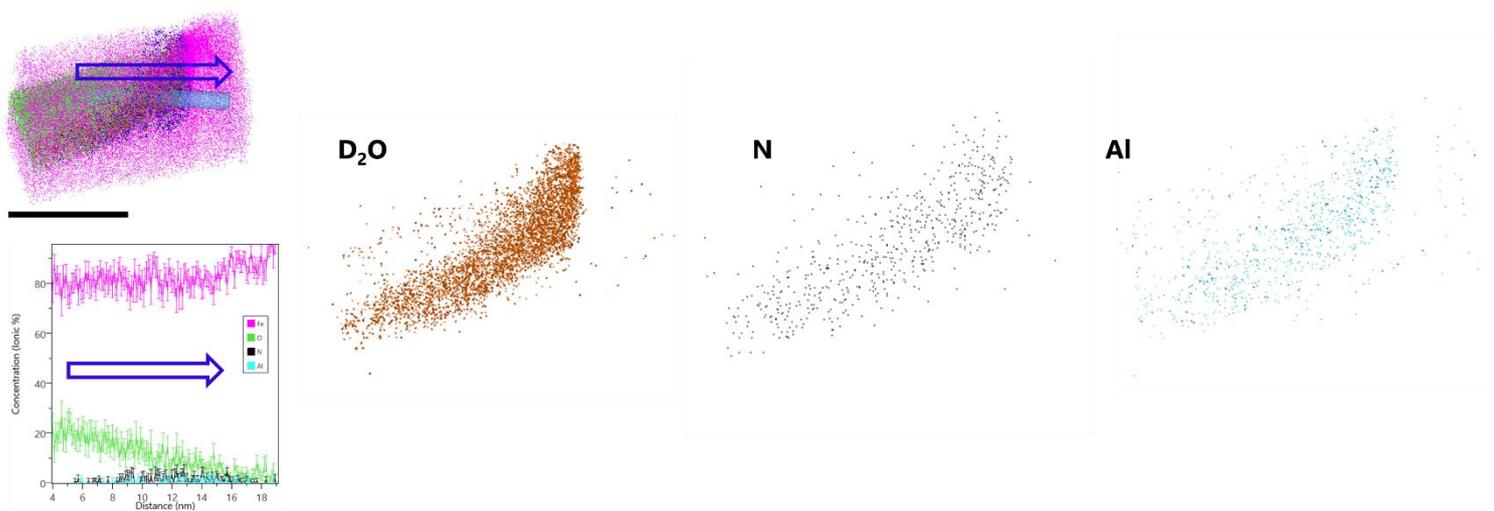

**Figure S10.** Local enrichment of impurities such as, N and Al at Ca-D$_2$O clusters formed after 10 seconds of reduction.

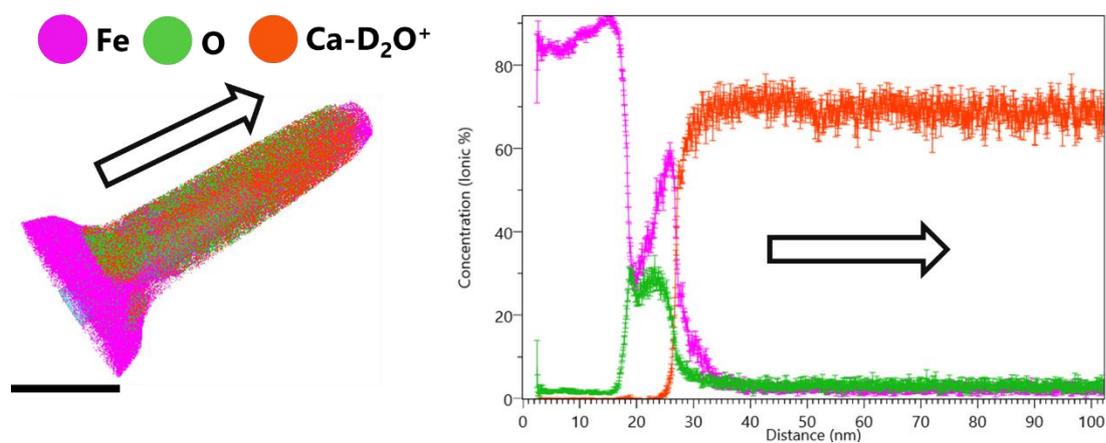

**Figure S11.** Growth of water-metal droplets after 60s of reduction. Scale bar is 20 nm.